\documentclass[aps,prl,twocolumn,showpacs,bibnotes]{revtex4}

\usepackage{graphicx}
\usepackage{amssymb}
\usepackage{bm}
\usepackage{here}
\newcommand{\ee}{\end{equation}}

\newcommand\pictc[5]{\begin{figure}
                   \centerline{
\includegraphics*[width=#1\columnwidth,height=0.7\textheight,keepaspectratio]{#3}}
               \protect\caption{\protect\label{fig:#4} #5}
                \end{figure}            }
\newcommand\pict[4][1]{\pictc{#1}{!tb}{#2}{#3}{#4}}
\newcommand\rpict[1]{\ref{fig:#1}}
\newcommand\leqt[1]{\protect\label{eq:#1}}
\newcommand\reqtn[1]{\ref{eq:#1}}
\newcommand\reqt[1]{(\reqtn{#1})}\newcounter{Fig}

\begin{document}

\begin{sloppy}

\title{Asymmetric vortex solitons in nonlinear periodic lattices}

\author{Tristram J. Alexander}
\author{Andrey A. Sukhorukov}
\author{Yuri S. Kivshar}
\affiliation{Nonlinear Physics Group and Center for Ultra-high
bandwidth Devices for Optical Systems (CUDOS), Research School of
Physical Sciences and Engineering, Australian National University,
Canberra ACT 0200, Australia}

\begin{abstract}
We reveal the existence of asymmetric vortex solitons in ideally
symmetric periodic lattices, and show how such nonlinear localized
structures describing elementary circular flows can be analyzed
systematically using the energy-balance relations. We present the
examples of {\em rhomboid}, {\em rectangular}, and {\em
triangular} vortex solitons on a square lattice, and also describe
novel coherent states where the populations of clockwise and
anti-clockwise vortex modes change periodically due to a
nonlinearity-induced momentum exchange through the lattice.
Asymmetric vortex solitons are expected to exist in different
nonlinear lattice systems including optically-induced photonic
lattices, nonlinear photonic crystals, and Bose-Einstein
condensates in optical lattices.
\end{abstract}

\pacs{
     42.65.Jx, %Beam trapping, self-focusing and defocusing; self-phase modulation
    42.65.Tg, %Optical solitons; nonlinear guided waves
      03.75.Lm, % Tunneling, Josephson effect, Bose-Einstein condensates
                % in periodic potentials, solitons, vortices
                % and topological excitations
     42.70.Qs %Photonic bandgap materials
     }\maketitle

Vortices are fundamental objects which 
appear in many branches of
physics~\cite{Pismen:1999:VorticesNonlinear}. In optics, vortices
are associated with phase dislocations (or phase singularities)
carried by diffracting optical
beams~\cite{Soskin:2001-219:ProgressOptics+}, and they share many
common properties with the vortices observed in superfluids and
Bose-Einstein condensates (BEC)~\cite{Madison:2000-806:PRL+}.
Vortices may emerge spontaneously or can be generated by various
experimental techniques, and they are important objects for
fundamental studies also having many practical applications.
Nonlinear self-action effects are inherent to many physical
systems, and they support the existence of {\em localized
vortex-like structures} in the form of vortex solitons when the
flow around the singularity does not change the density
distribution, the property naturally observed for axially
symmetric structures in homogeneous media [Fig.~\rpict{flow}(a)].

Periodic lattices, such as photonic structures for laser beams or
optical lattices for atomic BECs, allow for a strong modification
of the wave propagation, which also depends on the energy density.
Recently, it was demonstrated experimentally that photonic
lattices can support stable off-site (small radius) and on-site
(larger radius) vortices on a square lattice~\cite{Neshev+}. The
square-like profiles of such vortices resemble a homogeneous
vortex modulated by the underlying periodic structure
[Fig.~\rpict{flow}(b)]. Vortex solitons describe elementary
circular energy flows, and therefore their properties are
intrinsically linked to the wave transport mechanisms in the
underlying periodic lattice. In this Letter, we demonstrate that
this connection is highly nontrivial due to an interplay between
nonlinearity and periodicity. We predict analytically and confirm
by numerical simulations that even ideally symmetric periodic
structures can support robust {\em asymmetric vortex solitons},
and our approach allows for a systematic study of such novel types
of singular states. We find that some symmetries are always
allowed, whereas other configurations, such as a triangular vortex
shown in Fig.~\rpict{flow}(c), may exist under certain conditions
derived from a balance of the energy flows. Additionally, we
predict the existence of fully coherent states in the form of
vortices which exhibit {\em a charge flipping}, i.e. a periodic
reversal of the energy flow. This effect can occur for
larger-radius vortices, uncovering a key difference between the
on-site and off-site vortex states.

We consider the nonlinear propagation of an optical beam in a
two-dimensional periodic lattice~\cite{Christodoulides:2003-817:NAT} described by the dimensionless nonlinear equation,
\begin{equation} \leqt{nls}
   i \frac{\partial \Psi}{\partial z}
   + D \nabla_{\perp} \Psi + {V}( x, y) \Psi
   - {\cal G}( x, y, |\Psi|^2) \Psi
   = 0,
\end{equation}
where $ \nabla_{\perp}$ stands for the transverse Laplacian,
$\Psi(x,y,z)$ is the complex field amplitude, $x,y$ are the
transverse coordinates, $z$ is the propagation coordinate, $D$ is
the diffraction (or dispersion) coefficient. Function $V$ defines
a periodic potential of the two-dimensional lattice, and the
function ${\cal G}$ characterizes a nonlinear response. Similar
mathematical models appear for describing the self-action effects
in nonlinear photonic crystals~\cite{Mingaleev:2001-5474:PRL}, and
the nonlinear dynamics of atomic BEC in optical
lattices~\cite{Ostrovskaya:2003-160407:PRL+}.

\pict{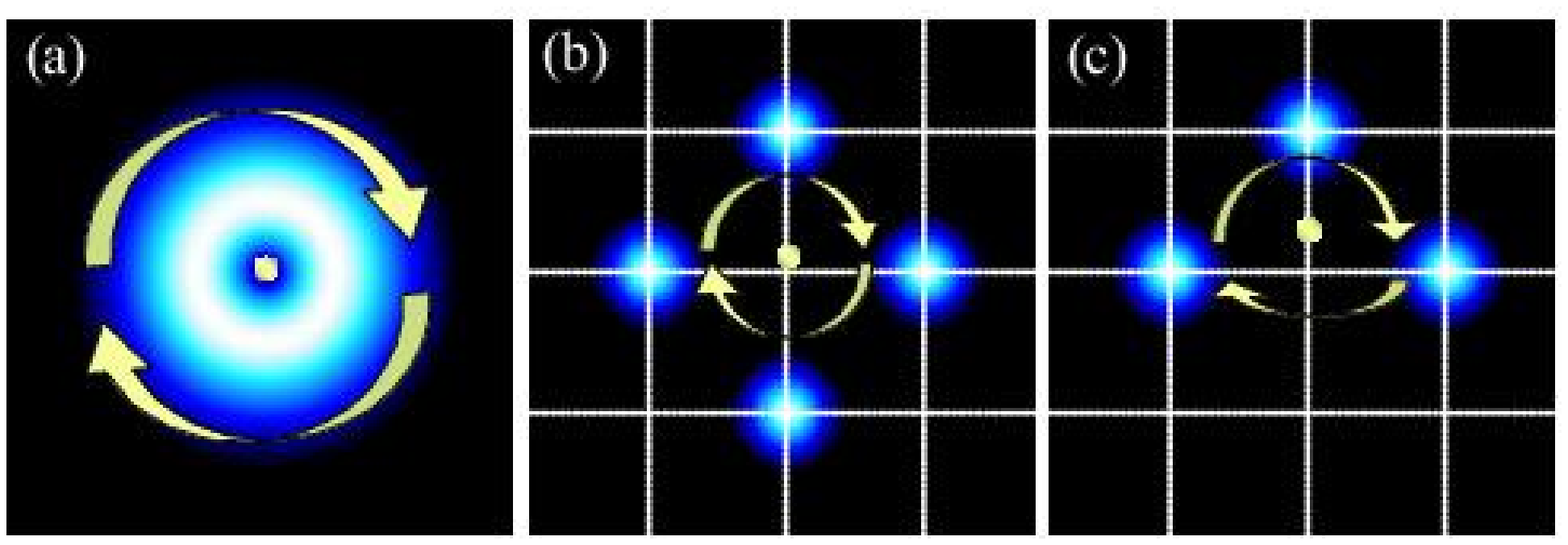}{flow}{Schematic of (a)~conventional vortex in a
homogeneous medium, (b,c)~symmetric and asymmetric vortex solitons
in a periodic lattice. Arrows indicate the direction of the phase
change and associated energy flow around the screw dislocation
(phase singularity) marked at the vortex center. }

Nonlinearity in Eq.~\reqt{nls} can compensate for the
diffraction-induced beam spreading in the transverse dimensions
leading to the formation of stationary structures in the form of
spatial solitons, $\Psi(x,y,z) = \psi(x,y) e^{i \beta z}$, where
$\psi(x,y)$ is the soliton envelope, and $\beta$ is a nonlinear
shift of the propagation constant, {\em the soliton parameter}.
Periodic lattices can modify dramatically the soliton properties,
as was demonstrated for waveguide arrays and dynamically-induced
periodic photonic lattices~\cite{Christodoulides:2003-817:NAT}. In
particular, periodic lattices can stabilize and support
propagation of {\em discrete vortex
solitons}~\cite{Johansson:1998-115:PD, Malomed:2001-26601:PRE,
Kevrekidis:2002-16609:PRE, Yang:2003-2094:OL+} which have recently
been observed in experiment~\cite{Neshev+}.

In order to analyze the vortex-like structures in a periodic
potential, we write the field envelope in the form,  $\psi(x,y) =
|\psi(x,y)|\exp[i \varphi(x,y)]$, and assume that the accumulation
of the phase $\varphi$ around a singular point (at $\psi=0$) is
$2\pi M$, where $M$ is an integer topological charge. We
consider spatially localized structures in the form of vortex-like
bright solitons with the envelopes vanishing at infinity. Such
structures can exist when the soliton parameter $\beta$ is placed
inside a gap of the linear Floquet-Bloch spectrum of the periodic
structure~\cite{Mingaleev:2001-5474:PRL,
Ostrovskaya:2003-160407:PRL+}.

The profiles of four-lobe {\em symmetric vortex solitons}
discussed so far in both the theory~\cite{Johansson:1998-115:PD,
Malomed:2001-26601:PRE, Kevrekidis:2002-16609:PRE,
Yang:2003-2094:OL+} and experiment~\cite{Neshev+} resemble closely a
ring-like structure of the soliton
clusters~\cite{Desyatnikov:2002-53901:PRL} in homogeneous media.
Then, we look for novel {\em vortex solitons of arbitrary symmetry} as a superposition of a number of the fundamental solitons,
\begin{equation} \leqt{cluster}
 \Psi(x,y;z) = \sum_{n=0}^{N-1} A_n(z) \psi_s(x-x_n, y-y_n) e^{i \beta z},
\end{equation}
where $\psi_s$ are the profiles of the individual fundamental
solitons, $n = 0,\ldots,N-1$, $N$ is the total number of solitons,
$(x_n, y_n)$ are the soliton positions, and $A_n$ is the 
scaling coefficient defining the phase of the $n$-th soliton and variation of its amplitude due to interaction with other solitons. In contrast to the case of a homogeneous medium~\cite{Desyatnikov:2002-53901:PRL}, the
positions of the individual solitons are fixed by the lattice
potential, provided the lattice potential is sufficiently strong.
In order to determine the soliton amplitudes, we present
Eq.~\reqt{nls} in the Hamiltonian form, $i d \Psi/d z = \delta
{\cal H}(x,y,\Psi,\Psi^\ast) / \delta \Psi^\ast$, and derive,
after substituting Eq.~\reqt{cluster} into the full Hamiltonian,
the reduced Hamiltonian $H_s(x,y,A_n,A_n^\ast)$ . The resulting
amplitude equations can be written in the form,
\begin{equation} \leqt{DNLS}
   i \frac{d A_n}{d z} = \frac{\delta H_s}{\delta A_n^\ast} =
     - \sum_{m=0}^{N-1} c_{n m} A_m - G( |A_n|^2 ) A_n - F_n,
\end{equation}
where $c_{n m} \equiv c_{m n} = C_{n m} / C_{n n}$, $C_{n m} =
\int\int \psi_s(x-x_n, y-y_n) \psi_s^\ast(x-x_m, y-y_m) dx dy$ are
the coupling coefficients, $G$ is the effective local
nonlinearity, and $F_n$ defines the nonlinear coupling terms such
as $\sim A_{m_1} A_{m_2}^\ast A_n$. We note that the
approximation~\reqt{cluster} is valid when $c_{n \ne m}, G, F/A_n
\sim \varepsilon \ll 1$. As was
demonstrated earlier~\cite{MacKay:1994-1623:NLN, Aubry:1997-201:PD}, under such conditions the amplitudes $A_n$ are only
slightly perturbed,
and we can seek stationary solutions of Eq.~\reqt{DNLS} corresponding to vortex solitons by means of the perturbation theory: $A_n = [1 + O(\varepsilon)] \exp[ i\varphi_n + i O(\varepsilon)]$. In Eq.~\reqt{DNLS}, the nonlinear coupling terms are proportional to the forth-order overlap
integrals and, therefore, $F_n \ll \sum_m c_{n m} A_m$. Then, in
the first order we obtain a general constraint on the soliton
phases $\varphi_n$,
\begin{equation} \leqt{balance}
  \sum_{m=0}^{N-1} c_{n m} \sin( \varphi_m - \varphi_n ) = 0 .
\end{equation}
In the sum~\reqt{balance}, each term defines the energy flow
between the solitons with numbers $n$ and $m$, so that the
equations~\reqt{balance} represent a condition for {\em a balance
of the energy flows} which is required for stable propagation of a
soliton cluster and the vortex-soliton formation. These conditions
are satisfied trivially when all the solitons are in- or
out-of-phase. We note that Eq.~\reqt{DNLS} with $F_n \equiv 0$ has
the form of {\em a discrete self-trapping equation}, which appears
in different physical contexts~\cite{Eilbeck:1985-318:PD,
Eilbeck:2003-44:Proc+}. However, nontrivial solutions of
Eqs.~\reqt{balance} corresponding to the vortex-like soliton
clusters have been analyzed only for symmetric configurations, and
even then some important solutions have been overlooked, as we
demonstrate below. Moreover, we show that the existence domains of
asymmetric vortex-like solutions are highly nontrivial, due to
specific properties of the coupling coefficients calculated for
realistic periodic lattices.

In order to provide a direct link to the recent experimental
results~\cite{Neshev+}, first we apply our
general analytical approach to describe the vortex solitons
in two-dimensional optically-induced lattices created in
a photorefractive crystal. Then, the diffraction coefficient in
Eq.~\reqt{nls} is defined as $D = z_0 \lambda / (4 \pi n_0
x_0^2)$, where $x_0$ and $z_0$ are the characteristic length-scales in
the transverse and longitudinal spatial dimensions, respectively,
$n_0$ is the average medium refractive index, and $\lambda$ is the
vacuum wavelength. The lattice potential and nonlinear beam
self-action effect are both due to the photorefractive screening
nonlinearity,
\begin{equation} \leqt{photorefr}
  {\cal G} - V = \gamma \left\{I_b + I_0
       \sin^2\left( \frac{\pi x}{d} \right)
       \sin^2\left( \frac{\pi y}{d} \right)
       +  |\Psi|^2 \right\}^{-1} ,
\end{equation}
where $\gamma$ is proportional to the external bias field,
$I_b$ is the dark irradiance, and $I_0$ is the intensity of
interfering beams that induce a square lattice with the period $d$ (see
details in Refs.~\cite{Efremidis:2003-213906:PRL, Neshev+}).

\pict{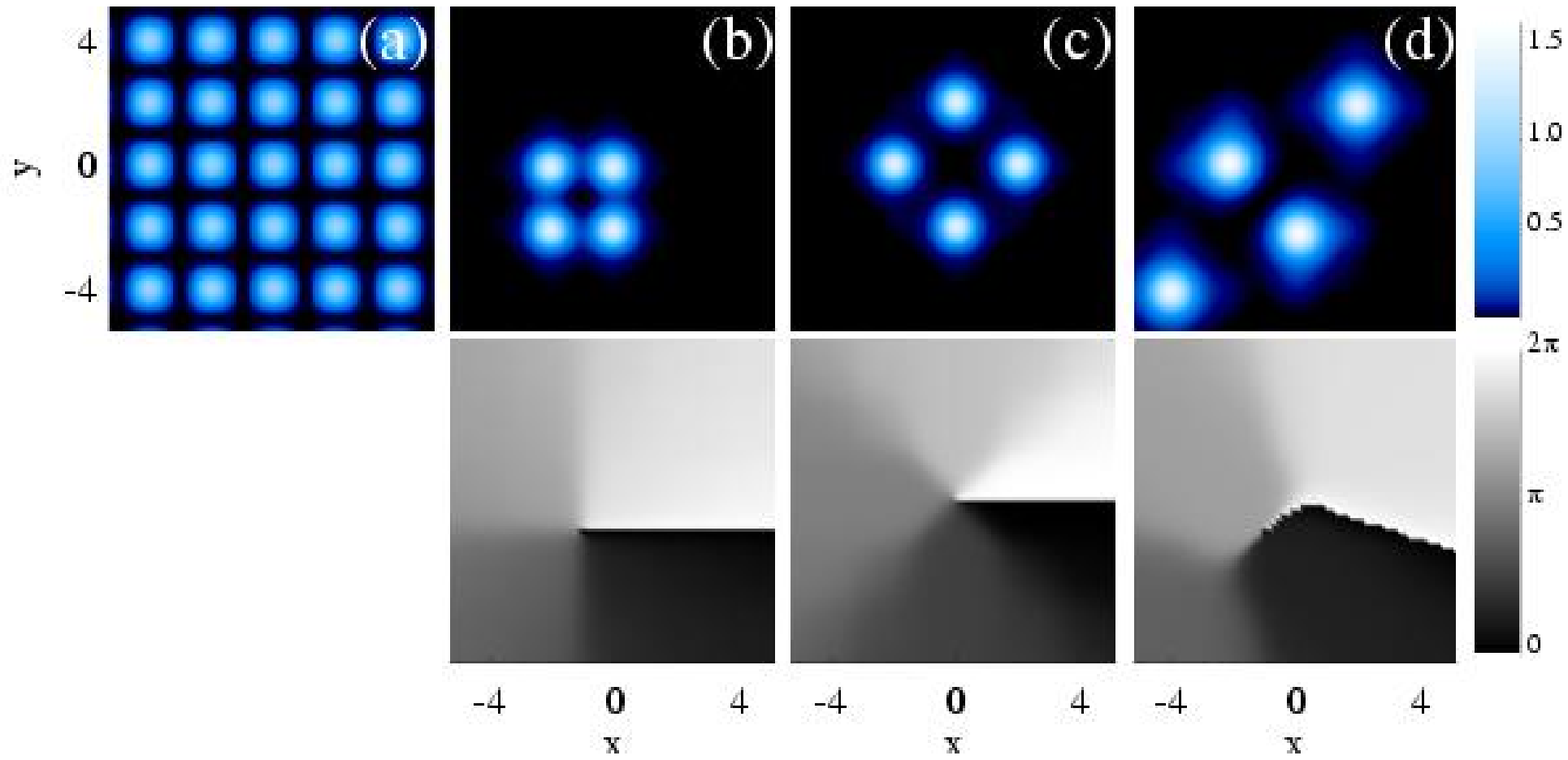}{vortex4}{Examples of the vortex-type soliton
structures with various symmetries in a square lattice potential
shown in (a). (b,c) Off-site and on-site symmetric square vortex solitons ($\gamma=8$, $\beta=5.5$); (d)~rhomboid configuration ($\gamma=5$, $\beta=3.5$) with a topological charge $+1$. Shown are the intensity profiles (top) and phase structures (bottom).}

We use nonlinear localized (soliton) solutions obtained by using
the ansatz~\reqt{cluster} as initial conditions for a numerical
relaxation algorithm applied to the
model~\reqt{nls},\reqt{photorefr}. We start by considering a {\em
four-soliton vortex} on a square lattice
[Fig.~\rpict{vortex4}(a)]. The previously reported solutions
[Figs.~\rpict{vortex4}(b,c)] have both the reflection and 90$^o$
rotational symmetries, similar to the underlying lattice.
Accordingly, $c_{0 1} = c_{1 2} = c_{2 3} = c_{3 0}$ and $c_{0 2}
= c_{1 3}$, and Eqs.~\reqt{balance} have a nontrivial-phase
solution with $\varphi_n = n \pi/2$ corresponding to a {\em
charge-one symmetric vortex}. However, our analytical approach
allows us to predict more general, {\em reduced-symmetry vortex
solitons} when the 90$^o$ rotational symmetry is removed. In
particular, we find a rhomboid configuration that has a vortex
charge with $\varphi_n = n \pi/2$ [Fig.~\rpict{vortex4}(d)]. We
found that rhomboid vortices are remarkably robust suggesting that
their generation in experiment can be possible by using
elliptically shaped singular beams.

Most remarkably, for both square and rhomboid configurations, we
find that the balance equations~\reqt{balance} admit more general
exact solutions, which were overlooked in the earlier
studies~\cite{Eilbeck:1985-318:PD} of Eq.~\reqt{DNLS}, namely
$\varphi_0 = \varphi_2 - \pi$ and $\varphi_1 = \varphi_3 - \pi$,
where {\em the phase difference $\varphi_1-\varphi_0$ is
arbitrary}. These novel solutions describe a family of vortex
solitons having the same intensity profile but different phase
structure. Due to such a degeneracy, a small change in the
amplitude of two opposing solitons can initiate a slow variation
of the free phase, $\varphi_1 -\varphi_0 \simeq \kappa z$; this
regime corresponds to {\em a periodic flipping of the vortex
charge}. Although the general equations~\reqt{DNLS} are only
satisfied when $\varphi_1-\varphi_0 = 0, \pi/2, \ldots$, the
charge-flipping effect can be induced by a finite perturbation
when the the nonlinear coupling terms ($F_n$) are small. Indeed,
we find that a closely packed vortex-like state shown in
Fig.~\rpict{vortex4}(b) is resistant to the charge flipping
effect, similar to vortices in single-well
potentials~\cite{Garcia-Ripoll:2001-140403:PRL}, whereas the
vortex shown in Fig.~\rpict{vortex4}(c) can exhibit the charge
flipping after increasing the amplitudes of two opposing solitons
by 7\%, as shown in Fig.~\rpict{flip}. The bottom plot
clearly illustrates that the solitons strongly exchange energy,
however the flows always remain balanced. These are {\em novel
coherent states}, where the populations of clockwise and
anti-clockwise rotational modes change periodically due to
nonlinearity-induced momentum exchange through the lattice.

\pict{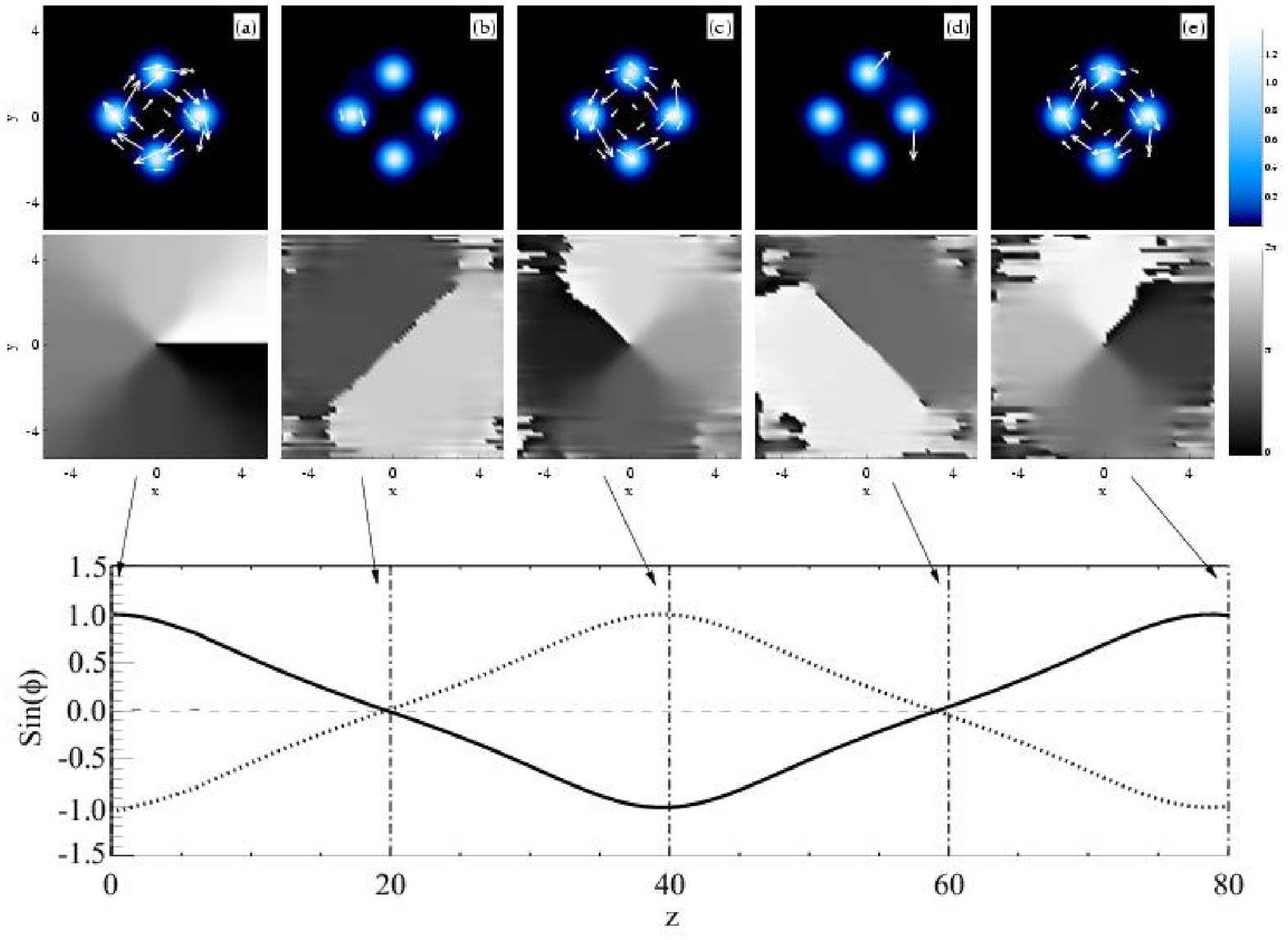}{flip}{Charge flipping effect for a square vortex of
Fig.~\rpict{vortex4}(c)  induced by a 7\% increase of the
amplitudes of two opposite solitons. Top: Snapshots at increasing
propagation distances showing the unchanged intensity profile.
Arrows show the energy flow.  Bottom: energy flows between the
solitons characterized by sinusoidal functions of the phase
differences between the opposite (dashed line) and neighboring
(solid and dotted lines) solitons according to
Eq.~\reqt{balance}.}

\pict[0.85]{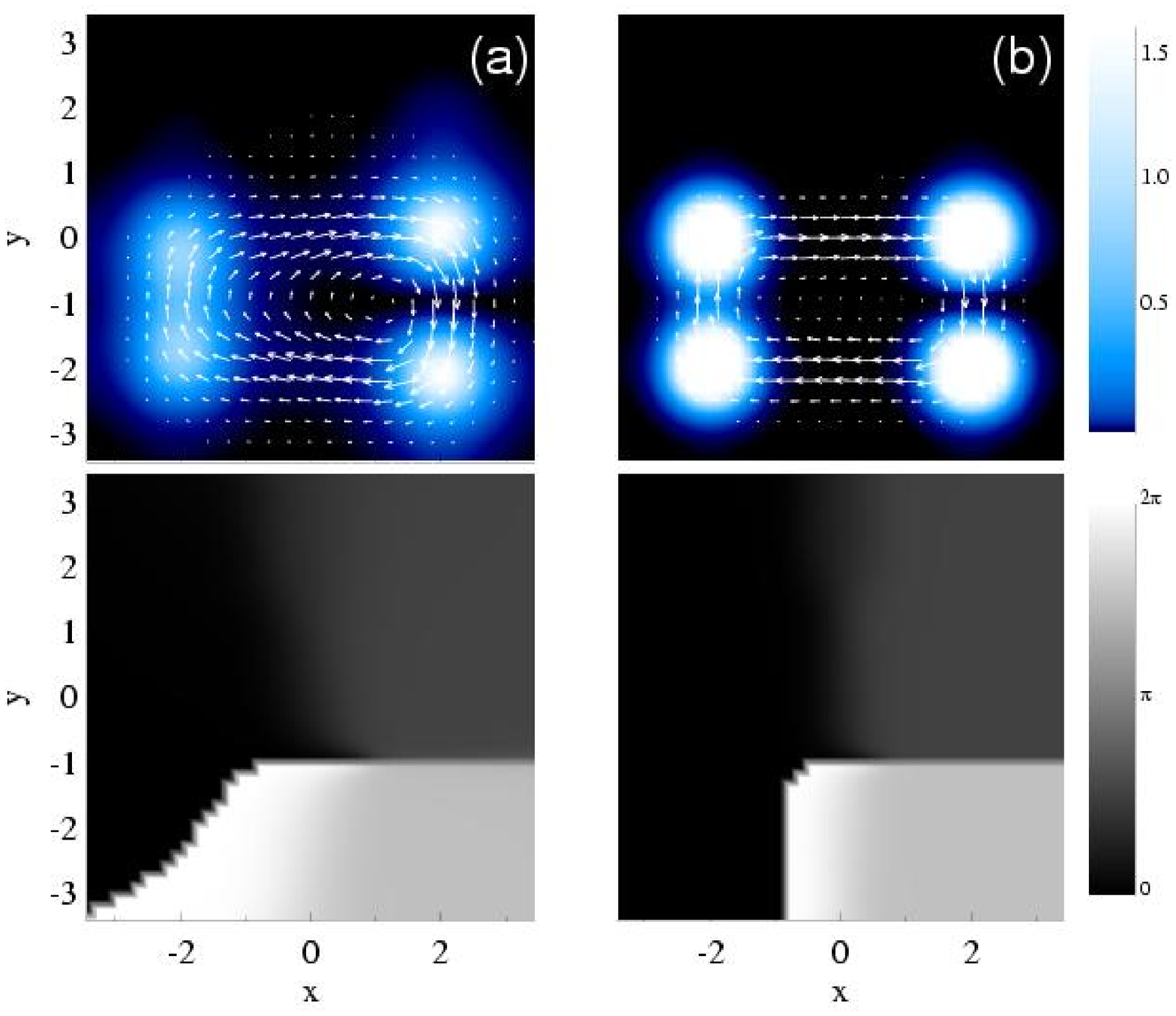}{rectangle}{Examples of asymmetric vortex solitons
with rectangle configurations. Top: intensity profiles with arrow
indicating the energy flow; Bottom:  phase structures. }

We now consider another example, vortices with rectangular
arrangement of the fundamental solitons. In this case, $c_{0 1} =
c_{2 3}$, $c_{0 3} = c_{1 2}$, and $c_{0 2} = c_{1 3}$. With no
loss of generality, we assign the soliton numbers such that $c_{0
1} \le c_{0 3}$, and then nontrivial solutions of
Eqs.~\reqt{balance} are: $\varphi_1-\varphi_0=\cos^{-1}(-c_{0
2}/c_{0 3})$, $\varphi_2-\varphi_0=\cos^{-1}(-c_{0 1}/c_{0 3})$,
and $\varphi_3+\varphi_0=\varphi_1+\varphi_2$. Since an inverse
cosine function has two branches, there exist four solutions
corresponding to two pairs of positively and negatively charged
vortices with different positions of singularity. We find that the
singularity is always shifted away from the center of rectangle
along its longer dimension, as illustrated in
Figs.~\rpict{rectangle}(a,b). This happens due to a highly
asymmetric phase structure of the vortex, which in turn can lead
to deformations of the vortex intensity profile resulting in a
trapezoid-like shape shown in Fig.~\rpict{rectangle}(a).

\pict{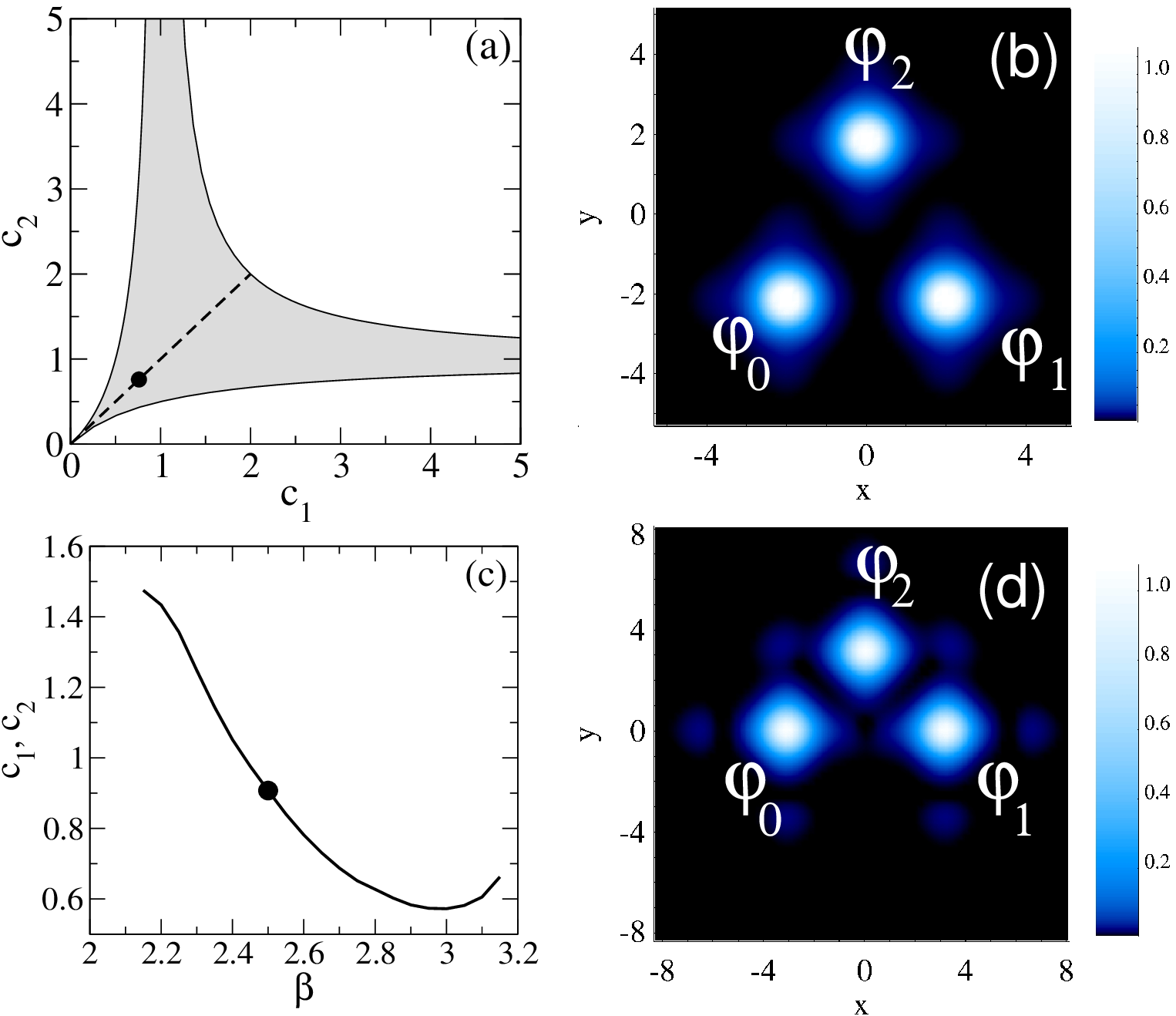}{triangle}{ (a)~Existence region in the plane of
normalized coupling coefficients for the three-soliton vortex;
(b)~Example of a vortex soliton with the phase configuration
$\varphi_0 = 0$, $\varphi_0 \simeq  0.981 \pi$, $\varphi_2 \simeq
-0.5097 \pi$, supported by a self-focusing photorefractive
screening nonlinearity ($\gamma=8$, $I_0=1$, $D=1$, $\beta=5.5$);
(c)~Dependence of the coupling coefficients on the soliton
propagation constant $\beta$ inside the band-gap for a configuration show in (d);
(d)~Example of a vortex soliton with the phase configuration
$\varphi_0=0$, $\varphi_1 \simeq 0.624 \pi$, $\varphi_2 \simeq
-0.625 \pi$, in repulsive atomic condensate ($\gamma=1$,
$V_0=2.5$, $D=1$) corresponding to marked points in plots (a,c)
($\beta = 2.5$). }

Finally, we mention another remarkable example of asymmetric
vortex solitons in the form of {\em a triangular structure}. We
find that Eqs.~\reqt{balance} possess nontrivial solutions for
$N=3$ only when the normalized coupling coefficients, $c_1 = c_{1
2} / c_{0 1}$ and $c_2 = c_{0 2} / c_{0 1}$, satisfy the following
conditions,
   $(1 + |c_1|^{-1})^{-1} < |c_2| < |1 - |c_1|^{-1}|^{-1}$,
and we show the existence domain in
Fig.~\rpict{triangle}(a). It follows that highly
asymmetric triangular vortices are not allowed. On the other hand,
isosceles configurations are always possible if the soliton base
is sufficiently narrow, so that condition $c_1=c_2 < 2$ is
satisfied. We show an example of such vortex soliton in
Fig.~\rpict{triangle}(b).

In order to underline the generality and broad applicability of
our results, we also search for asymmetric vortex solitons in
repulsive atomic Bose-Einstein condensates loaded onto a
two-dimensional optical lattice. In the mean-field approximation,
this system is described by the Gross-Pitaevskii
equation~\reqt{nls} with
\begin{equation}
  V = V_0 [\sin( \pi x/d ) + \sin( \pi y/d )]; \quad
  {\cal G} = \gamma |\Psi|^2 ,
\end{equation}
where $\gamma > 0$, and $V_0$ is the depth of an optical-lattice
potential. Repulsive nonlinearity can lead to wave localization in
the form of gap solitons~\cite{Ostrovskaya:2003-160407:PRL+}, which
exhibit effective long-range interaction. This allows, in
particular, the existence of triangular vortices with a wider base
[Fig.~\rpict{triangle}(d)] compared to the conventional solitons
in the self-focusing regime [cf.~Fig.~\rpict{triangle}(b)]. We
find that such {\em triangular gap vortices} exist throughout the whole
band-gap since $c_1=c_2 < 2$ [Fig.~\rpict{triangle}(c)], except
for the immediate vicinity of the spectrum band edges where our
approximation~\reqt{cluster} is not applicable.

In addition, we have verified that all vortex-like structures
discussed above are created by {\em strongly coupled solitons}.
This is an important issue, since it may be formally possible to
construct various composite states of almost decoupled solitons,
which exchange the energy at an extremely slow rate, so that
Eqs.~\reqt{balance} are nearly satisfied. In contrast, in the
cases discussed above there is a strong internal energy flow
within the phase-locked and charge-flipping vortices, as seen in
Fig.~\rpict{vortex4} to Fig.~\rpict{triangle}.

In conclusion, we have revealed that periodic lattices can support
different types of robust asymmetric vortex-like nonlinear
localized structures. Such vortices resemble the soliton clusters
trapped by the lattice, but they are associated with a non-trivial
power flow. We have presented the examples of novel 
vortex solitons on a square lattice, and other solutions 
can be obtained and analyzed using the general energy-balance relations. 
We believe our findings will initiate the experimental
efforts to observe such vortices in optically-induced photonic
structures, Bose-Einstein condensates in optical lattices,
photonic crystals, and photonic crystal fibers.

\end{sloppy}
\end{document}